\documentclass[prl,superscriptaddress,twocolumn,amsmath]{revtex4-1}

\usepackage{bm}
\usepackage{graphicx}
\usepackage[usenames,dvipsnames,svgnames,table]{xcolor}
\usepackage[colorlinks=true,linkcolor=RoyalBlue,citecolor=RoyalBlue]{hyperref}

\newcommand{\eps}{\varepsilon}

\newcommand{\ket}[1]{\lvert #1 \rangle}

\begin{document}

\title{Impact of novel electron-phonon coupling mechanisms on valley physics in two-dimensional materials}

\author{Wen-Yu Shan}
\affiliation{Department of Physics, School of Physics and Materials Science, Guangzhou University, Guangzhou 510006, China}

\date{\today}

\begin{abstract}
  We systematically study the impact of various electron-acoustic-phonon coupling mechanisms
  on valley physics in two-dimensional materials. In the static
  strain limit, we find that Dirac cone tilt and deformation potential have analogous valley Hall response since they fall into the same universality class of pseudospin structure. However, such argument fails for the coupling mechanism with position-dependent Fermi velocity. For the isotropic case, a significant valley Hall effect occurs near charge neutrality similar to the bond-length change, whereas for the anisotropic case, the geometric valley transport is suppressed, akin to the deformation potential. Gap opening mechanism by nonuniform strain is found to totally inhibit the valley Hall transport, even if the dynamics of strains are introduced. By varying gate voltage, a tunable phonon-assisted valley Hall response can be realized, which paves a way toward rich phenomena and new functionalities of valley
  acoustoelectronics.

\end{abstract}

\maketitle

$\textit{Introduction}$.---Electron-phonon coupling is known to play a pivotal role in determining the physical properties of solids, such as superconducting, optical, transport phenomena, etc. Particularly, recent advances in topological materials trigger new interest in understanding the interplay between electron-phonon coupling and topological physics~\cite{saha2014, saha2015, moller2017}. Due to the inherent topology and high tunability, two-dimensional (2D) materials provide unprecedented opportunities to investigate such interplay. There are growing evidences in 2D materials~\cite{engels2014, couto2014, amorim2016, wagner2019} which underline the significance of coupling between electrons and phonons, or their classical description, i.e., random strain fluctuations~\cite{couto2014, amorim2016}. Up to now, most transport studies of electron-phonon coupling concentrate on the longitudinal direction or spin relaxation~\cite{suzuura2002, mariani2008, castro2010, ochoa2011, ochoa2013}, however, little attention has been paid to the transverse transport~\cite{gorini2015, xiao20191, xiao20192}, e.g., anomalous, thermal, spin or valley Hall effect. Experimentally, transverse transport measurements have been realized in several 2D materials~\cite{gorbachev2014, sui2015, shimazaki2015, wu2019, hung2019}, whose results suggest an important role of electron-phonon coupling for realistic samples~\cite{shan2019}. This may open up new avenues for phonon-controlled technological applications of electrons, like valley acoustoelectronics~\cite{kalameitsev2019}.

New physics also stems from the novel electron-phonon coupling mechanisms in 2D materials. For example, the pseudogauge potential due to the bond-length change in monolayer graphene~\cite{vozmediano2010} exhibits distinct behaviors from the usual deformation potential, including the Landau-level formation~\cite{guinea2010, levy2010}, quantum spin Hall effect~\cite{cazalilla2014}, etc. Intriguingly, recent work unveils that the pseudogauge-type electron-phonon coupling significantly enhances the valley Hall response near charge neutrality for Dirac electrons~\cite{shan2019}, as compared to the deformation type. Such difference is physically related to the distinct universality classes of pseudospin structure for electron-phonon coupling~\cite{yang2011}: orthogonal (deformation-type), unitary (magnetic type) and symplectic (gauge type). Nevertheless, there are more exotic types beyond the standard classification: Dirac cone tilt (DT)~\cite{goerbig2008, choi2010, kobayashi2007}, isotropic velocity renormalization (IR)~\cite{winkler2010, dejuan2012}, anisotropic velocity renormalization (AR)~\cite{winkler2010, dejuan2012, dejuan2013} and gap opening by nonuniform strain (GO)~\cite{manes2013}. The novelties of these types lie in the nontrivial interplay between electron's orbital motion, strain tensor $u_{\alpha\beta}$ and (pseudo) spin structure, whose effect on transport phenomena remains largely unexplored. All these types contain up to the first derivatives of the strain tensor or electron fields as the standard classes, implying that any type may be the dominant electron-phonon coupling mechanisms in realistic 2D materials~\cite{amorim2016, winkler2010}. Therefore studying the impact of these new types would be desirable for better understanding and manipulation of electron-phonon coupling in 2D materials.

In this paper, we explore the impact of all leading-order electron-phonon coupling mechanisms allowed by symmetries on valley physics in graphene systems. In the static strain limit, we find that the DT type and the deformation potential have analogous valley Hall response since they belong to the same universality class of pseudospin structure. The IR type, similar to the gauge-type electron-phonon coupling, significantly enhances the valley Hall effect near charge neutrality for Dirac electrons, whereas the AR type, akin to the deformation potential, suppresses such geometric transport. The GO type totally inhibits the valley Hall transport, even if the dynamics of strains are introduced. Electron-phonon coupling is further investigated to demonstrate these features. Our study refreshes our knowledge on both electron-phonon coupling and valley physics, and paves the way toward valley acoustoelectronics for 2D materials.

$\textit{Model and strains}$.---We begin with the effective Hamiltonian
\begin{equation} \label{ham} 
H_0 = \hbar v\bm k \cdot \bm\sigma + \Delta\sigma_z \;,
\end{equation}
which describes the low-energy electron dynamics in one of the Dirac
valleys in gapped graphene.  Here $\bm k = (k_x, k_y)$ is the two-dimensional wave vector, $\bm\sigma$
represents the sublattice indices.  The other
valley can be obtained by performing a time-reversal operation on
$H_0$.  The energy dispersion is given by
$\eps_{c,v} = \pm \eps_{\bm k} = \pm (\hbar^2v^2k^2 + \Delta^2)^{1/2}$, where the subscript $c$ and $v$ label the conduction and valence
bands, respectively. Without loss of generality, we suppose the Fermi level lies in the conduction band. The corresponding eigenstate is given by $\ket{u_{\bm k}^c} = [\cos(\theta_{\bm k}/2), \sin(\theta_{\bm k}/2)e^{i\phi_{\bm k}}]^T$, where the angular variables $\theta_{\bm k}= \cos^{-1}(\Delta/\eps_{\bm k})$ and $\phi_{\bm k} = \tan^{-1}(k_y/k_x)$. In the low temperature limit, electronic transport is basically a Fermi surface property, and a single parameter $\theta_F=\theta|_{|\bm k|=k_F}$ can be used to characterize the transport behaviors, where $k_F$ is the Fermi wave vector.

For the phononic part, it is convenient to first treat it classically, that is, by using the language of strains~\cite{landau1959}. In the long-wavelength limit, static strain vector $\bm u(\bm r)$ due to the atomic deviations from equilibrium positions in solids, has in-plane and out-of-plane (flexural) modes. For either mode, we require that the spatial average of strain vanishes, i.e., $\langle \bm u(\bm r) \rangle_{av}=0$. This suggests that at the mean-field level, there is no strain-induced modification to band structure or topology, in contrast to the usual analysis of strain patterns~\cite{cazalilla2014}. Such requirement can be reconciled by considering the random distribution of strains, namely, random strain fluctuations~\cite{couto2014, amorim2016}, which occur as a natural consequence of thermally excited phonons. To attack the new issue, we need to go beyond the mean-field level, and concern the fluctuation effect of strains. For clarity, we adopt a detailed form of strain vector  $\bm u (\bm r)=\sum_{\bm q}\bm u(\bm q)e^{i\bm q\cdot\bm r}$, where
\begin{equation}\label{form1} 
\bm u(\bm q) = F_{\bm q}\frac{\bm q}{|\bm q|^{\eta}} \;
\end{equation}
in a single strained region and $F_{\bm q}$ is arbitrary $\bm q$-dependent function. Such form originates from the long-wavelength acoustic phonon modes of graphene systems. For random strains, fluctuations from different regions are correlated and contribute to the transport properties through their correlation function $\langle F_{\bm q}F_{-\bm q}\rangle_{av}=C|\bm q|^{\epsilon}$ by the harmonic approximation of phonons. Here $\eta=4$, $\epsilon=2$ ($\eta=3$, $\epsilon=0$) denote the out-of-plane (in-plane) mode of graphene~\cite{couto2014}, physically due to thermal fluctuation effect or random forces from substrates. This suggests that two-dimensional materials are not really flat. $C$ is a material-dependent parameter.

$\textit{Electron-strain (-fluctuation) coupling}$.---The description of electron-strain coupling relies on a rank-2 strain tensor, $u_{\alpha\beta}=\frac{1}{2}[\partial_{\alpha}u_{\beta}+\partial_{\beta}u_{\alpha}+\partial_{\alpha}h\partial_{\beta}h]$, where $\bm u=(u_x,u_y)$ ($h$) is the in-plane (out-of-plane) displacement for a particular phonon mode. Usually up to the leading order of $u_{\alpha\beta}$, the coupling with strain reads
\begin{equation} 
\begin{split}\label{e-strain0}
\hat{V}(\bm r) &= g_{d}\sum_{\alpha}u_{\alpha\alpha} + g_{s}\sum_{\alpha}u_{\alpha\alpha}\sigma_z + g_{b}\bm a\cdot\bm \sigma\;
\end{split}
\end{equation}
for 2D hexagonal crystals like graphene and transition-metal dichalcogenides~\cite{cazalilla2014}. $\bm a=(u_{xx}-u_{yy},-2u_{xy})$. Microscopically, $g_d$, $g_s$ and $g_b$ term originate from uniform, staggered deformation and bond-length change of lattices, respectively. In the language of random matrix~\cite{dyson1962}, they correspond to the orthogonal (deformation-type), unitary (pseudomagnetic-type) and symplectic (gauge-type) classes, respectively. Due to the distinct pseudospin structure, transport behaviors of Dirac electrons may be qualitatively different among all three classes, including magnetoconductivity~\cite{shan2012}, anomalous~\cite{yang2011} and valley Hall effect~\cite{shan2019}.

Conventionally, these classes cover all possibilities for disordered (pseudo) spin-1/2 system. However, apart from the pseudospin structure, the electron's orbital motion $\hat{\bm p}=-i\hbar\nabla_{\bm r}$, when combined with the strain tensor $u_{\alpha\beta}$, may lead to new paradigms. These unconventional situations can be understood by noting that the underlying hexagonal lattice puts strong constraints on the form of electron-strain coupling. In this sense, one can follow the symmetry analysis~\cite{winkler2010,manes2013} to find all possible types up to the first derivative of $u_{\alpha\beta}$. For instance, in graphene with $D_{6h}$ group, new types beyond $g_d$, $g_b$ (without $g_s$) in Eq. (\ref{e-strain0}) are found as Dirac cone tilt (DT)
\begin{equation}\label{e-strain1}
\hat{V}(\bm r) = g_{DT}\bm a\cdot\hat{\bm p}+H.c.,
\end{equation}
isotropic velocity renormalization (IR)
\begin{equation}\label{e-strain2} 
\hat{V}(\bm r) = g_{IR} \sum_{\alpha}u_{\alpha\alpha}\hat{\bm p}\cdot\bm\sigma + H.c. \;,
\end{equation}
anisotropic velocity renormalization (AR)
\begin{equation}\label{e-strain3} 
\hat{V}(\bm r) = g_{AR} \sum_{\alpha,\beta}u_{\alpha\beta}\hat{p}_{\alpha}\sigma_{\beta} + H.c. \;
\end{equation}
and gap opening by nonuniform strain (GO)
\begin{equation}\label{e-strain4} 
\hat{V}(\bm r) = g_{GO} [\nabla\times\bm a]_z\sigma_z  \;.
\end{equation}
$H.c.$ denotes the Hermitian conjugation. The DT type is pronounced under uniaxial strain along the zigzag direction~\cite{goerbig2008, choi2010}, and also in two-dimensional organic conductors~\cite{kobayashi2007}. The IR and AR type contribute to the isotropic and anisotropic position-dependent Fermi velocity, evidenced by the scanning tunnelling spectroscopy of graphene grown on SiO$_2$ thermal oxide~\cite{luican2011}, Rh foil~\cite{yan2013}, boron nitride~\cite{jang2014}. The GO type describes a Zeeman coupling of pseudospin to the pseudomagnetic field $\bm B=\nabla\times\bm a$, with an energy approximately 7 meV~\cite{manes2013}. Compared to the $g_d$, $g_b$ term, these new types show a clear nontrivial $\hat{\bm p}$ dependence, whose effect on transport behaviors is worthy of investigation. Moreover, since the averaged strain vanishes as mentioned above, the electron-strain coupling actually reduces to an exotic \textit{electron-strain-fluctuation} coupling, which may introduce new intriguing features in the transport behaviors. In the following, we mainly focus on the valley Hall physics.

$\textit{Coordinate shift}$.---To unveil the valley Hall effect, let us first evaluate an essential quantity, the coordinate shift $\delta\bm r_{\bm k'\bm k}$. It describes a shift of electron wave packet scattered from the state with average momentum $\bm k$ into the one with $\bm k'$ by random strain fluctuations. The disorder-averaged gauge-invariant form of $\delta\bm r_{\bm k'\bm k}$ follows~\cite{sinitsyn2006}
\begin{equation} 
\delta\bm r_{\bm k'\bm k} = \bm A^c_{\bm k'} -\bm A^c_{\bm k} -\langle\hat{\bm D}_{\bm k',\bm k}\arg[V^c_{\bm k'\bm k}]\rangle_{av},
\end{equation}
where $\bm A^c_{\bm k}=\langle u^c_{\bm k}|i\nabla_{\bm k}u^c_{\bm k}\rangle$ is the Berry connection of electrons. The Born amplitude between Bloch states $|\psi^c_{\bm k}\rangle$ and $|\psi^c_{\bm k'}\rangle$ is given by $V^c_{\bm k'\bm k}=\langle\psi^c_{\bm k'}|\hat{V}(\bm r)|\psi^c_{\bm k}\rangle$, where $|\psi^c_{\bm k}\rangle=e^{i\bm k\cdot\bm r}|u^c_{\bm k}\rangle$. Note that for new types of $\hat{V}(\bm r)$, additional contributions may arise when individually acting on $\langle\psi^c_{\bm k'}|$ and $|\psi^c_{\bm k}\rangle$. The argument $\arg[V^c_{\bm k'\bm k}]=-i\ln[V^c_{\bm k'\bm k}/|V^c_{\bm k'\bm k}|]$ and $\hat{\bm D}_{\bm k',\bm k}=\nabla_{\bm k'}+\nabla_{\bm k}$. An order-of-magnitude estimate of $x$-component coordinate shift $(\delta\bm r_{\bm k'\bm k})_x$ is given in Table ~\ref{tab:type_coupling}, where different types of electron-strain-fluctuation coupling are summarized. For comparison, conventional types are also listed. Since our interest is mainly around the charge neutrality point of Dirac electrons, a parameter $\sin\theta_F\ll1$, proportional to $\sqrt{n}$ (the carrier density $n$), is useful to characterize $(\delta\bm r_{\bm k'\bm k})_x$.

\renewcommand\arraystretch{1.5}

\begin{table}[bp]
\caption{ Order-of-magnitude estimate of coordinate shift $(\delta\bm r_{\bm k'\bm k})_x$ (in terms of $\sin\theta_F=\hbar vk_F/\epsilon_F\ll1$ and phonon energy $\omega_{\bm q}\ll\epsilon_F$) and explicit form of valley Hall conductivity $\sigma_{xy}^v$ (in units of $\frac{2e^2}{h}$) near charge neutrality in the low temperature limit for different types of electron-strain-fluctuation coupling. Static (dynamical) case corresponds to $\omega_{\bm q}=|\epsilon_{\bm k}-\epsilon_{\bm k'}|=0$ ($\neq0$). $\bm q=\bm k'-\bm k$ is the momentum transfer of electrons. $\epsilon_F$ is the Fermi energy and $k_F$ is the Fermi wave vector. $\chi$ is a cutoff-related factor~\cite{supple}. $l$ ($t$) labels longitudinal (transverse) in-plane phonon modes. $\surd$ ($\times$) refers to enhanced (suppressed) valley Hall response.}
\label{tab:type_coupling}%
\begin{ruledtabular}
\begin{tabular}{*4c} 
Types of & Phonon  & $(\delta\bm r_{\bm k'\bm k})_x$  & $\sigma_{xy}^v[\frac{2e^2}{h}]$ \\ 
electron-strain-fluctuation  & modes & $\propto$ & $=$  \\
coupling   &  &  &   \\ \hline
DT type $(\times)$ & $l$, $t$ & $\sin\theta_F$  & $-\frac{\sin^2\theta_F\cos\theta_F}{1-\frac{1}{4}\sin^2\theta_F}$ \\ 
IR type $(\surd)$ & $l$ & $[\sin\theta_F]^{-1}$  & $\cos\theta_F$ \\
(dynamical) AR type $(\times)$ & $l$, $t$  & $\sin\theta_F$ & $-\chi\sin^2\theta_F\cos\theta_F$ \\
(static) GO type $(\times)$ & $l$, $t$ &  $0$ & $0$  \\
(dynamical) GO type $(\times)$ & $l$, $t$ & $\omega_{\bm q}\sin\theta_F$  & $-\frac{\omega_{\bm q}}{2\epsilon_F}\frac{\sin^2\theta_F\cos\theta_F}{1-\frac{1}{4}\sin^2\theta_F}$  \\ \hline
deformation type $(\times)$ & $l$  & $\sin\theta_F$ & $-\frac{\sin^2\theta_F\cos\theta_F}{1-\frac{1}{2}\sin^2\theta_F}$  \\
pseudomagnetic type $(\times)$ & $l$ & $0$  & $0$  \\
gauge type $(\surd)$ & $l$, $t$ & $[\sin\theta_F]^{-1}$  & $\cos\theta_F$  \\
\end{tabular}
\end{ruledtabular}
\end{table}

For the DT type, $(\delta\bm r_{\bm k'\bm k})_{x}\propto\sin\theta_F$ is vanishingly small, similar to the deformation type. This can be understood by noting that there is no internal pseudospin structure for either type of electron-strain-fluctuation coupling. As a result, $\hat{V}(\bm r)$ does not induce a pseudospin-flipping transition between initial and final states $|\psi^c_{\bm k/\bm k'}\rangle$ (both are almost spin up), thus the Born amplitude $V^c_{\bm k'\bm k}$ becomes finite. This finite Born amplitude then leads to a tiny coordinate shift as $(\delta\bm r_{\bm k'\bm k})_x\propto\sin\theta_F/\langle|V^c_{\bm k'\bm k}|^2\rangle_{av}$.

For the IR type, $(\delta\bm r_{\bm k'\bm k})_x$ diverges as $[\sin\theta_F]^{-1}$, similar to the gauge type. The reason is that either $\hat{\bm p}\cdot\bm\sigma$ in Eq. (\ref{e-strain2}) or $\bm a\cdot\bm\sigma$ in Eq. (\ref{e-strain0}) is composed of pseudospin-flipping operators, whose Born amplitude between the same-spin states $|\psi^c_{\bm k/\bm k'}\rangle$ is vanishingly small as $V^c_{\bm k'\bm k}\propto\sin\theta_F$. Moreover, the coefficient $g_{IR}\sum_{\alpha}u_{\alpha\alpha}$, recognized as isotropic position-dependent Fermi velocity, has no influence on the final result. This agrees with a conventional interpretation based on universality classification of pseudospin structures. Nevertheless, such argument fails for the AR type. Despite being constructed by spin-flipping operators $\sigma_{x,y}$, the dynamical AR type shows $\sim\sin\theta_F$ behavior of $(\delta\bm r_{\bm k'\bm k})_x$, akin to the deformation type. Here dynamical (static) case corresponds to nonzero (zero) energy transfer of electrons during scattering, or equivalently, phonon energy $\omega_{\bm q}=|\epsilon_{\bm k}-\epsilon_{\bm k'}|\neq0$ ($=0$), where $\bm q=\bm k'-\bm k$ is the momentum transfer. Since the static AR type is ill-defined, the dynamical type becomes dominant. Intriguingly, an explicit form of $(\delta\bm r_{\bm k'\bm k})_x$ is independent of phonon energy $\omega_{\bm q}$, demonstrating the validity and stability of the dynamical type.

The static GO type contributes to zero coordinate shift, even when the Fermi level is away from the charge neutrality point. This result is the same as the pseudomagnetic type, since their pseudospin share a common $\sigma_z$ structure. Furthermore, when we go beyond the paradigm to the dynamical case, a phonon energy-dependent $(\delta\bm r_{\bm k'\bm k})_x$ occurs.

$\textit{Valley Hall conductivity}$.---Once $(\delta\bm r_{\bm k'\bm k})_x$ is derived, one can follow the Boltzmann equation~\cite{sinitsyn2006} to evaluate the valley Hall conductivity $\sigma_{xy}^v$. The explicit form of $\sigma_{xy}^v$ in the low temperature limit is shown in Table ~\ref{tab:type_coupling}, whose magnitude is linked to $(\delta\bm r_{\bm k'\bm k})_x$ by a scaling relation $\sigma_{xy}^v\propto\sin\theta_F(\delta\bm r_{\bm k'\bm k})_x$. The properties of $\sigma_{xy}^v$ are thus directly inherited from $(\delta\bm r_{\bm k'\bm k})_x$. As a result, we note that the IR type leads to a near-quantized valley Hall conductivity, as found for the gauge type~\cite{shan2019}. The DT and dynamical AR type tend to suppress $\sigma_{xy}^v$ near the charge neutrality point. The intriguing difference of transport responses by IR and AR type is related to the distinct structure of electron-phonon coupling, which forbids the scattering of electrons for AR type when the strains are static. This suggests that the transport due to AR type is purely a dynamical effect, whereas IR type is a static effect. Furthermore, for the dynamical GO type, given the prefactors $\omega_{\bm q}/\epsilon_F$ and $\sin^2\theta_F$ in $\sigma_{xy}^v$, valley Hall response is totally inhibited. Based on these analysis, particularly on the IR and dynamical AR type, we point out that a conventional universality classification of (pseudo) spin structure for electron-phonon coupling is not enough to characterize the transverse transport behaviors of Dirac electrons, such as the valley Hall effect. These constitute the main results of our paper. Additionally, $\sigma_{xy}^v$ is independent of specific choice of phonon modes (different $\eta$ and $\epsilon$ in Eq. (\ref{form1})), manifesting the generality of our results. For the gauge type, the near-quantized $\sigma_{xy}^v$ can be also understood based on a pseudomagnetic-field picture, i.e., $\bm B\propto f(|\bm r|)y(3x^2-y^2)\hat{z}$, where $f(|\bm r|)$ is an angle-independent function. Such field agrees with the three-fold rotational symmetry of graphene. By contrast, this picture fails for IR and AR types since their form cannot be treated as Peierls substitution of Dirac systems. This further highlights the difference of IR and AR types from the conventional electron-phonon coupling.

\begin{figure}[t]
\centering \includegraphics[width=0.48\textwidth]{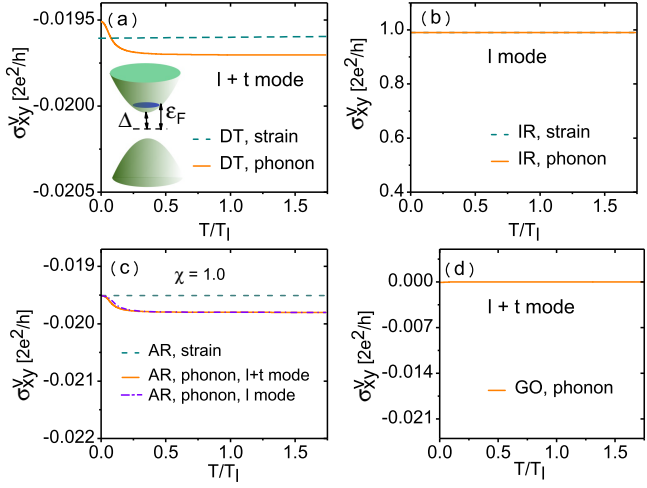}
\caption{ Temperature dependence of side-jump valley Hall conductivity (in units of $2e^2/h$) via (a) DT-type, (b) IR-type, (c) AR-type and (d) GO-type coupling. Yellow solid (green dashed) curve corresponds to the electron-phonon (electron-strain-fluctuation) coupling in the low-temperature limit. In (a), (c) and (d), mixed longitudinal ($l$) and transverse ($t$) acoustic modes are discussed; in (b), a single longitudinal mode is analyzed. Parameters: $\Delta=50$ meV, $\epsilon_F=1.01\Delta$, $v=1\times10^6$ m/s, $c_l=2.10\times10^4$ m/s, $c_t=1.38\times10^4$ m/s~\cite{amorim2016}. The Bloch-Gruneisen temperature $T_l=3.5$ K, $T_t=2.3$ K. The fitting cutoff-related factor $\chi=1.0$~\cite{supple}.}
\label{fig:ephonon}
\end{figure}

$\textit{Electron-phonon coupling}$.---Now we provide a quantum treatment of the electron-strain-fluctuation coupling, that is, the electron-phonon coupling. First, it is convenient to decompose the strain vector $\bm u(\bm q)$ into longitudinal $(l)$ and transverse $(t)$ modes: $u_{\bm q}^{l}=\bm u\cdot\bm q/q$ and $u_{\bm q}^{t}=\bm u\cdot(\hat{z}\times\bm q/q)$. The phonon field in the long-wavelength limit reads $u_{\bm q}^{\nu}=i\sqrt{\frac{\hbar}{2M\omega_{\bm q}^{\nu}}}(d_{\bm q}^{\nu}+d^{\nu+}_{-\bm q})$, where $d_{\bm q}$ ($d_{\bm q}^+$) refers to the annihilation (creation) operator of acoustic phonon modes $\nu=l$, $t$. $M$ is the oscillator mass and $\omega_{\bm q}^{\nu}=c_{\nu}q$ is the phonon dispersion, where $c_{l/t}$ is the longitudinal (transverse) sound speed. By introducing the phonon field $u_{\bm q}^{\nu}$, the coupling form $\hat{V}(\bm r)$ can be reorganized into $l$ or $t$ mode~\cite{supple}. In contrast to the classical treatment on pure longitudinal ($l$) mode in Eq. (\ref{form1}), here both $l$ and $t$ modes are considered, as summarized in Table ~\ref{tab:type_coupling}. We employ a semiclassical approach to deal with the phonon-assisted valley Hall response, by which the deformation type has been discussed~\cite{xiao20191}.

New types of electron-phonon coupling are systematically analyzed in our paper~\cite{supple}. Numerical results are shown in Fig.~\ref{fig:ephonon}, where temperature dependences of side-jump valley Hall conductivity are studied for low electron-doped samples, i.e., $\Delta<\epsilon_F\ll2\Delta$. To gain insights, we plot the temperature behaviors of $\sigma_{xy}^v$ due to the electron-strain-fluctuation coupling and electron-phonon coupling, respectively. For the IR and AR type (see Fig.~\ref{fig:ephonon} (b)-(c)), different treatment of couplings yields consistent results in the low temperature limit, $T\ll T_l$, $T_t$ (with the Bloch-Gruneisen temperature $T_{l/t}=2c_{l/t}k_F/k_B$), whereas for the DT type (in Fig.~\ref{fig:ephonon} (a)), there is a minor discrepancy. The consistency of order of magnitude of $\sigma_{xy}^v$ justifies the treatment of electron-strain-fluctuation coupling. Nevertheless, the AR- and GO-type electron-strain-fluctuation coupling (see Fig.~\ref{fig:ephonon} (c)-(d)) fails to provide exact results due to a lack of knowledge of cutoff factor $\chi$ or averaged phonon energy $\omega_{\bm q}$. In this sense, in Fig.~\ref{fig:ephonon} (c), $\chi=1.0$ is derived by fitting $\sigma_{xy}^v$ from the electron-phonon coupling in the zero-temperature limit. In Fig.~\ref{fig:ephonon} (d), $\sigma_{xy}^v$ from the electron-phonon coupling is vanishingly small, about $10^{-5}e^2/h$, as a result of small ratio $c_{l/t}/v$ and $\sin^3\theta_F$ as indicated in Table~\ref{tab:type_coupling}. This verifies the physics that the GO type totally inhibits the valley Hall response.

For the DT, AR and GO type, a mixture of $l$ and $t$ mode is discussed, whereas for the IR type, only $l$ mode is plotted. The difference between IR and AR type does not arise from the occurrence of $t$ mode, which can be understood from the behavior of pure $l$ mode in Fig.~\ref{fig:ephonon} (c). Note that all temperature dependences originate from the Bose distribution functions of phonons, which is appropriate for $T/T_l<1$. When further increasing the temperature, i.e., $T\gg T_l$, the small discrepancy between the two couplings should be smeared out by introducing the Fermi distribution functions of electrons, which is out of the scope of our work.

$\textit{Discussion}$.---Here, we have systematically studied the effect of electron-acoustic-phonon coupling on valley physics in graphene systems. All possible types of coupling allowed by lattice symmetries are individually investigated. Accordingly, the dominant type of coupling in realistic samples can be determined by the distinct valley Hall responses, in combination with other experimental probes, such as scanning tunnelling spectroscopy~\cite{yan2013,jang2014}, Raman spectroscopy~\cite{chakraborty2012} or magnetophonon resonance~\cite{kumaravadivel2019}. 
By varying gate voltage (thus $\theta_F$), a tunable phonon-assisted valley Hall response can be realized, which constitutes a new type of valley acoustoelectronics, in contrast to the conventional one driven by surface acoustic waves~\cite{kalameitsev2019}.

A few remarks are in order. First, our framework is in stark contrast with the usual studies of electron-strain coupling which rely on the band renormalization~\cite{saha2014, saha2015, moller2017}. The reason is that near the band edge, a self-energy correction to the band structure is negligible when compared to the large band gap. As a result, any geometric or topological modification to valley Hall response is not expected. Instead, we focus on the fluctuation and correlation effect of strains, which up to now have received far less attention. This understanding of our paper underlines the importance of spatial dynamics of strains. On the other hand, for the temporal dynamics of strains, we describe an evolution of valley Hall response from static to dynamical situation, where qualitative differences are found.

Second, the phonons we concern here are the low-energy acoustic $\Gamma$ phonons, where there is no Berry curvature for phonon bands due to their degeneracy. This means the induced valley Hall response is not a valley phonon Hall effect stemming from pure phonon topology~\cite{zhang2015}. Rather, it is a consequence of electron topology and particularly the exotic electron-phonon coupling. Similarly, anomalous and spin Hall effect driven by the coupling between electrons and fluctuating magnets~\cite{ishizuka2018,kato2019,okamoto2019} have attracted recent interest.

Finally, our results are not restricted to graphene systems, but can also be applied for other two-dimensional materials, such as monolayer transition-metal dichalcogenide MoS$_2$~\cite{cazalilla2014,rostami2015}. It is also appealing to extend our theoretical framework to twisted bilayer graphene~\cite{ochoa2019,lian2019}, where the occurrence of such new types of electron-phonon coupling may lead to exotic valley- and phonon-related phenomena. 

This work is supported by the National Natural Science Foundation of China (NSFC, Grant No. 11904062). We also acknowledge the support of a startup grant from Guangzhou University.


\begin{thebibliography}{51}%
\makeatletter
\providecommand \@ifxundefined [1]{%
 \@ifx{#1\undefined}
}%
\providecommand \@ifnum [1]{%
 \ifnum #1\expandafter \@firstoftwo
 \else \expandafter \@secondoftwo
 \fi
}%
\providecommand \@ifx [1]{%
 \ifx #1\expandafter \@firstoftwo
 \else \expandafter \@secondoftwo
 \fi
}%
\providecommand \natexlab [1]{#1}%
\providecommand \enquote  [1]{``#1''}%
\providecommand \bibnamefont  [1]{#1}%
\providecommand \bibfnamefont [1]{#1}%
\providecommand \citenamefont [1]{#1}%
\providecommand \href@noop [0]{\@secondoftwo}%
\providecommand \href [0]{\begingroup \@sanitize@url \@href}%
\providecommand \@href[1]{\@@startlink{#1}\@@href}%
\providecommand \@@href[1]{\endgroup#1\@@endlink}%
\providecommand \@sanitize@url [0]{\catcode `\\12\catcode `\$12\catcode
  `\&12\catcode `\#12\catcode `\^12\catcode `\_12\catcode `\%12\relax}%
\providecommand \@@startlink[1]{}%
\providecommand \@@endlink[0]{}%
\providecommand \url  [0]{\begingroup\@sanitize@url \@url }%
\providecommand \@url [1]{\endgroup\@href {#1}{\urlprefix }}%
\providecommand \urlprefix  [0]{URL }%
\providecommand \Eprint [0]{\href }%
\providecommand \doibase [0]{http://dx.doi.org/}%
\providecommand \selectlanguage [0]{\@gobble}%
\providecommand \bibinfo  [0]{\@secondoftwo}%
\providecommand \bibfield  [0]{\@secondoftwo}%
\providecommand \translation [1]{[#1]}%
\providecommand \BibitemOpen [0]{}%
\providecommand \bibitemStop [0]{}%
\providecommand \bibitemNoStop [0]{.\EOS\space}%
\providecommand \EOS [0]{\spacefactor3000\relax}%
\providecommand \BibitemShut  [1]{\csname bibitem#1\endcsname}%
\let\auto@bib@innerbib\@empty
\bibitem [{\citenamefont {Saha}\ and\ \citenamefont {Garate}(2014)}]{saha2014}%
  \BibitemOpen
  \bibfield  {author} {\bibinfo {author} {\bibfnamefont {K.}~\bibnamefont
  {Saha}}\ and\ \bibinfo {author} {\bibfnamefont {I.}~\bibnamefont {Garate}},\
  }\href {\doibase 10.1103/PhysRevB.89.205103} {\bibfield  {journal} {\bibinfo
  {journal} {Phys. Rev. B}\ }\textbf {\bibinfo {volume} {89}},\ \bibinfo
  {pages} {205103} (\bibinfo {year} {2014})}\BibitemShut {NoStop}%
\bibitem [{\citenamefont {Saha}\ \emph {et~al.}(2015)\citenamefont {Saha},
  \citenamefont {L\'egar\'e},\ and\ \citenamefont {Garate}}]{saha2015}%
  \BibitemOpen
  \bibfield  {author} {\bibinfo {author} {\bibfnamefont {K.}~\bibnamefont
  {Saha}}, \bibinfo {author} {\bibfnamefont {K.}~\bibnamefont {L\'egar\'e}}, \
  and\ \bibinfo {author} {\bibfnamefont {I.}~\bibnamefont {Garate}},\ }\href
  {\doibase 10.1103/PhysRevLett.115.176405} {\bibfield  {journal} {\bibinfo
  {journal} {Phys. Rev. Lett.}\ }\textbf {\bibinfo {volume} {115}},\ \bibinfo
  {pages} {176405} (\bibinfo {year} {2015})}\BibitemShut {NoStop}%
\bibitem [{\citenamefont {M\"{o}ller}\ \emph {et~al.}(2017)\citenamefont
  {M\"{o}ller}, \citenamefont {Sawatzky}, \citenamefont {Franz},\ and\
  \citenamefont {Berciu}}]{moller2017}%
  \BibitemOpen
  \bibfield  {author} {\bibinfo {author} {\bibfnamefont {M.~M.}\ \bibnamefont
  {M\"{o}ller}}, \bibinfo {author} {\bibfnamefont {G.~A.}\ \bibnamefont
  {Sawatzky}}, \bibinfo {author} {\bibfnamefont {M.}~\bibnamefont {Franz}}, \
  and\ \bibinfo {author} {\bibfnamefont {M.}~\bibnamefont {Berciu}},\ }\href
  {\doibase 10.1038/s41467-017-02442-y} {\bibfield  {journal} {\bibinfo
  {journal} {Nature Communications}\ }\textbf {\bibinfo {volume} {8}},\
  \bibinfo {pages} {2267} (\bibinfo {year} {2017})}\BibitemShut {NoStop}%
\bibitem [{\citenamefont {Engels}\ \emph {et~al.}(2014)\citenamefont {Engels},
  \citenamefont {Terr\'es}, \citenamefont {Epping}, \citenamefont {Khodkov},
  \citenamefont {Watanabe}, \citenamefont {Taniguchi}, \citenamefont
  {Beschoten},\ and\ \citenamefont {Stampfer}}]{engels2014}%
  \BibitemOpen
  \bibfield  {author} {\bibinfo {author} {\bibfnamefont {S.}~\bibnamefont
  {Engels}}, \bibinfo {author} {\bibfnamefont {B.}~\bibnamefont {Terr\'es}},
  \bibinfo {author} {\bibfnamefont {A.}~\bibnamefont {Epping}}, \bibinfo
  {author} {\bibfnamefont {T.}~\bibnamefont {Khodkov}}, \bibinfo {author}
  {\bibfnamefont {K.}~\bibnamefont {Watanabe}}, \bibinfo {author}
  {\bibfnamefont {T.}~\bibnamefont {Taniguchi}}, \bibinfo {author}
  {\bibfnamefont {B.}~\bibnamefont {Beschoten}}, \ and\ \bibinfo {author}
  {\bibfnamefont {C.}~\bibnamefont {Stampfer}},\ }\href {\doibase
  10.1103/PhysRevLett.113.126801} {\bibfield  {journal} {\bibinfo  {journal}
  {Phys. Rev. Lett.}\ }\textbf {\bibinfo {volume} {113}},\ \bibinfo {pages}
  {126801} (\bibinfo {year} {2014})}\BibitemShut {NoStop}%
\bibitem [{\citenamefont {Couto}\ \emph {et~al.}(2014)\citenamefont {Couto},
  \citenamefont {Costanzo}, \citenamefont {Engels}, \citenamefont {Ki},
  \citenamefont {Watanabe}, \citenamefont {Taniguchi}, \citenamefont
  {Stampfer}, \citenamefont {Guinea},\ and\ \citenamefont
  {Morpurgo}}]{couto2014}%
  \BibitemOpen
  \bibfield  {author} {\bibinfo {author} {\bibfnamefont {N.~J.~G.}\
  \bibnamefont {Couto}}, \bibinfo {author} {\bibfnamefont {D.}~\bibnamefont
  {Costanzo}}, \bibinfo {author} {\bibfnamefont {S.}~\bibnamefont {Engels}},
  \bibinfo {author} {\bibfnamefont {D.-K.}\ \bibnamefont {Ki}}, \bibinfo
  {author} {\bibfnamefont {K.}~\bibnamefont {Watanabe}}, \bibinfo {author}
  {\bibfnamefont {T.}~\bibnamefont {Taniguchi}}, \bibinfo {author}
  {\bibfnamefont {C.}~\bibnamefont {Stampfer}}, \bibinfo {author}
  {\bibfnamefont {F.}~\bibnamefont {Guinea}}, \ and\ \bibinfo {author}
  {\bibfnamefont {A.~F.}\ \bibnamefont {Morpurgo}},\ }\href {\doibase
  10.1103/PhysRevX.4.041019} {\bibfield  {journal} {\bibinfo  {journal} {Phys.
  Rev. X}\ }\textbf {\bibinfo {volume} {4}},\ \bibinfo {pages} {041019}
  (\bibinfo {year} {2014})}\BibitemShut {NoStop}%
\bibitem [{\citenamefont {{Amorim}}\ \emph {et~al.}(2016)\citenamefont
  {{Amorim}}, \citenamefont {{Cortijo}}, \citenamefont {{de Juan}},
  \citenamefont {{Grushin}}, \citenamefont {{Guinea}}, \citenamefont
  {{Guti{\'e}rrez-Rubio}}, \citenamefont {{Ochoa}}, \citenamefont {{Parente}},
  \citenamefont {{Rold{\'a}n}}, \citenamefont {{San-Jos{\'e}}}, \citenamefont
  {{Schiefele}}, \citenamefont {{Sturla}},\ and\ \citenamefont
  {{Vozmediano}}}]{amorim2016}%
  \BibitemOpen
  \bibfield  {author} {\bibinfo {author} {\bibfnamefont {B.}~\bibnamefont
  {{Amorim}}}, \bibinfo {author} {\bibfnamefont {A.}~\bibnamefont {{Cortijo}}},
  \bibinfo {author} {\bibfnamefont {F.}~\bibnamefont {{de Juan}}}, \bibinfo
  {author} {\bibfnamefont {A.~G.}\ \bibnamefont {{Grushin}}}, \bibinfo {author}
  {\bibfnamefont {F.}~\bibnamefont {{Guinea}}}, \bibinfo {author}
  {\bibfnamefont {A.}~\bibnamefont {{Guti{\'e}rrez-Rubio}}}, \bibinfo {author}
  {\bibfnamefont {H.}~\bibnamefont {{Ochoa}}}, \bibinfo {author} {\bibfnamefont
  {V.}~\bibnamefont {{Parente}}}, \bibinfo {author} {\bibfnamefont
  {R.}~\bibnamefont {{Rold{\'a}n}}}, \bibinfo {author} {\bibfnamefont
  {P.}~\bibnamefont {{San-Jos{\'e}}}}, \bibinfo {author} {\bibfnamefont
  {J.}~\bibnamefont {{Schiefele}}}, \bibinfo {author} {\bibfnamefont
  {M.}~\bibnamefont {{Sturla}}}, \ and\ \bibinfo {author} {\bibfnamefont
  {M.~A.~H.}\ \bibnamefont {{Vozmediano}}},\ }\href {\doibase
  10.1016/j.physrep.2015.12.006} {\bibfield  {journal} {\bibinfo  {journal}
  {Phys. Rep.}\ }\textbf {\bibinfo {volume} {617}},\ \bibinfo {pages} {1}
  (\bibinfo {year} {2016})}\BibitemShut {NoStop}%
\bibitem [{\citenamefont {Wagner}\ \emph {et~al.}(2020)\citenamefont {Wagner},
  \citenamefont {Nguyen},\ and\ \citenamefont {Simon}}]{wagner2019}%
  \BibitemOpen
  \bibfield  {author} {\bibinfo {author} {\bibfnamefont {G.}~\bibnamefont
  {Wagner}}, \bibinfo {author} {\bibfnamefont {D.~X.}\ \bibnamefont {Nguyen}},
  \ and\ \bibinfo {author} {\bibfnamefont {S.~H.}\ \bibnamefont {Simon}},\
  }\href {\doibase 10.1103/PhysRevLett.124.026601} {\bibfield  {journal}
  {\bibinfo  {journal} {Phys. Rev. Lett.}\ }\textbf {\bibinfo {volume} {124}},\
  \bibinfo {pages} {026601} (\bibinfo {year} {2020})}\BibitemShut {NoStop}%
\bibitem [{\citenamefont {Suzuura}\ and\ \citenamefont
  {Ando}(2002)}]{suzuura2002}%
  \BibitemOpen
  \bibfield  {author} {\bibinfo {author} {\bibfnamefont {H.}~\bibnamefont
  {Suzuura}}\ and\ \bibinfo {author} {\bibfnamefont {T.}~\bibnamefont {Ando}},\
  }\href {\doibase 10.1103/PhysRevB.65.235412} {\bibfield  {journal} {\bibinfo
  {journal} {Phys. Rev. B}\ }\textbf {\bibinfo {volume} {65}},\ \bibinfo
  {pages} {235412} (\bibinfo {year} {2002})}\BibitemShut {NoStop}%
\bibitem [{\citenamefont {Mariani}\ and\ \citenamefont {von
  Oppen}(2008)}]{mariani2008}%
  \BibitemOpen
  \bibfield  {author} {\bibinfo {author} {\bibfnamefont {E.}~\bibnamefont
  {Mariani}}\ and\ \bibinfo {author} {\bibfnamefont {F.}~\bibnamefont {von
  Oppen}},\ }\href {\doibase 10.1103/PhysRevLett.100.076801} {\bibfield
  {journal} {\bibinfo  {journal} {Phys. Rev. Lett.}\ }\textbf {\bibinfo
  {volume} {100}},\ \bibinfo {pages} {076801} (\bibinfo {year}
  {2008})}\BibitemShut {NoStop}%
\bibitem [{\citenamefont {Castro}\ \emph {et~al.}(2010)\citenamefont {Castro},
  \citenamefont {Ochoa}, \citenamefont {Katsnelson}, \citenamefont {Gorbachev},
  \citenamefont {Elias}, \citenamefont {Novoselov}, \citenamefont {Geim},\ and\
  \citenamefont {Guinea}}]{castro2010}%
  \BibitemOpen
  \bibfield  {author} {\bibinfo {author} {\bibfnamefont {E.~V.}\ \bibnamefont
  {Castro}}, \bibinfo {author} {\bibfnamefont {H.}~\bibnamefont {Ochoa}},
  \bibinfo {author} {\bibfnamefont {M.~I.}\ \bibnamefont {Katsnelson}},
  \bibinfo {author} {\bibfnamefont {R.~V.}\ \bibnamefont {Gorbachev}}, \bibinfo
  {author} {\bibfnamefont {D.~C.}\ \bibnamefont {Elias}}, \bibinfo {author}
  {\bibfnamefont {K.~S.}\ \bibnamefont {Novoselov}}, \bibinfo {author}
  {\bibfnamefont {A.~K.}\ \bibnamefont {Geim}}, \ and\ \bibinfo {author}
  {\bibfnamefont {F.}~\bibnamefont {Guinea}},\ }\href {\doibase
  10.1103/PhysRevLett.105.266601} {\bibfield  {journal} {\bibinfo  {journal}
  {Phys. Rev. Lett.}\ }\textbf {\bibinfo {volume} {105}},\ \bibinfo {pages}
  {266601} (\bibinfo {year} {2010})}\BibitemShut {NoStop}%
\bibitem [{\citenamefont {Ochoa}\ \emph {et~al.}(2011)\citenamefont {Ochoa},
  \citenamefont {Castro}, \citenamefont {Katsnelson},\ and\ \citenamefont
  {Guinea}}]{ochoa2011}%
  \BibitemOpen
  \bibfield  {author} {\bibinfo {author} {\bibfnamefont {H.}~\bibnamefont
  {Ochoa}}, \bibinfo {author} {\bibfnamefont {E.~V.}\ \bibnamefont {Castro}},
  \bibinfo {author} {\bibfnamefont {M.~I.}\ \bibnamefont {Katsnelson}}, \ and\
  \bibinfo {author} {\bibfnamefont {F.}~\bibnamefont {Guinea}},\ }\href
  {\doibase 10.1103/PhysRevB.83.235416} {\bibfield  {journal} {\bibinfo
  {journal} {Phys. Rev. B}\ }\textbf {\bibinfo {volume} {83}},\ \bibinfo
  {pages} {235416} (\bibinfo {year} {2011})}\BibitemShut {NoStop}%
\bibitem [{\citenamefont {Ochoa}\ \emph {et~al.}(2013)\citenamefont {Ochoa},
  \citenamefont {Guinea},\ and\ \citenamefont {Fal'ko}}]{ochoa2013}%
  \BibitemOpen
  \bibfield  {author} {\bibinfo {author} {\bibfnamefont {H.}~\bibnamefont
  {Ochoa}}, \bibinfo {author} {\bibfnamefont {F.}~\bibnamefont {Guinea}}, \
  and\ \bibinfo {author} {\bibfnamefont {V.~I.}\ \bibnamefont {Fal'ko}},\
  }\href {\doibase 10.1103/PhysRevB.88.195417} {\bibfield  {journal} {\bibinfo
  {journal} {Phys. Rev. B}\ }\textbf {\bibinfo {volume} {88}},\ \bibinfo
  {pages} {195417} (\bibinfo {year} {2013})}\BibitemShut {NoStop}%
\bibitem [{\citenamefont {Gorini}\ \emph {et~al.}(2015)\citenamefont {Gorini},
  \citenamefont {Eckern},\ and\ \citenamefont {Raimondi}}]{gorini2015}%
  \BibitemOpen
  \bibfield  {author} {\bibinfo {author} {\bibfnamefont {C.}~\bibnamefont
  {Gorini}}, \bibinfo {author} {\bibfnamefont {U.}~\bibnamefont {Eckern}}, \
  and\ \bibinfo {author} {\bibfnamefont {R.}~\bibnamefont {Raimondi}},\ }\href
  {\doibase 10.1103/PhysRevLett.115.076602} {\bibfield  {journal} {\bibinfo
  {journal} {Phys. Rev. Lett.}\ }\textbf {\bibinfo {volume} {115}},\ \bibinfo
  {pages} {076602} (\bibinfo {year} {2015})}\BibitemShut {NoStop}%
\bibitem [{\citenamefont {Xiao}\ \emph
  {et~al.}(2019{\natexlab{a}})\citenamefont {Xiao}, \citenamefont {Liu},
  \citenamefont {Xie}, \citenamefont {Yang},\ and\ \citenamefont
  {Niu}}]{xiao20191}%
  \BibitemOpen
  \bibfield  {author} {\bibinfo {author} {\bibfnamefont {C.}~\bibnamefont
  {Xiao}}, \bibinfo {author} {\bibfnamefont {Y.}~\bibnamefont {Liu}}, \bibinfo
  {author} {\bibfnamefont {M.}~\bibnamefont {Xie}}, \bibinfo {author}
  {\bibfnamefont {S.~A.}\ \bibnamefont {Yang}}, \ and\ \bibinfo {author}
  {\bibfnamefont {Q.}~\bibnamefont {Niu}},\ }\href {\doibase
  10.1103/PhysRevB.99.245418} {\bibfield  {journal} {\bibinfo  {journal} {Phys.
  Rev. B}\ }\textbf {\bibinfo {volume} {99}},\ \bibinfo {pages} {245418}
  (\bibinfo {year} {2019}{\natexlab{a}})}\BibitemShut {NoStop}%
\bibitem [{\citenamefont {Xiao}\ \emph
  {et~al.}(2019{\natexlab{b}})\citenamefont {Xiao}, \citenamefont {Liu},
  \citenamefont {Yuan}, \citenamefont {Yang},\ and\ \citenamefont
  {Niu}}]{xiao20192}%
  \BibitemOpen
  \bibfield  {author} {\bibinfo {author} {\bibfnamefont {C.}~\bibnamefont
  {Xiao}}, \bibinfo {author} {\bibfnamefont {Y.}~\bibnamefont {Liu}}, \bibinfo
  {author} {\bibfnamefont {Z.}~\bibnamefont {Yuan}}, \bibinfo {author}
  {\bibfnamefont {S.~A.}\ \bibnamefont {Yang}}, \ and\ \bibinfo {author}
  {\bibfnamefont {Q.}~\bibnamefont {Niu}},\ }\href {\doibase
  10.1103/PhysRevB.100.085425} {\bibfield  {journal} {\bibinfo  {journal}
  {Phys. Rev. B}\ }\textbf {\bibinfo {volume} {100}},\ \bibinfo {pages}
  {085425} (\bibinfo {year} {2019}{\natexlab{b}})}\BibitemShut {NoStop}%
\bibitem [{\citenamefont {Gorbachev}\ \emph {et~al.}(2014)\citenamefont
  {Gorbachev}, \citenamefont {Song}, \citenamefont {Yu}, \citenamefont
  {Kretinin}, \citenamefont {Withers}, \citenamefont {Cao}, \citenamefont
  {Mishchenko}, \citenamefont {Grigorieva}, \citenamefont {Novoselov},
  \citenamefont {Levitov},\ and\ \citenamefont {Geim}}]{gorbachev2014}%
  \BibitemOpen
  \bibfield  {author} {\bibinfo {author} {\bibfnamefont {R.~V.}\ \bibnamefont
  {Gorbachev}}, \bibinfo {author} {\bibfnamefont {J.~C.~W.}\ \bibnamefont
  {Song}}, \bibinfo {author} {\bibfnamefont {G.~L.}\ \bibnamefont {Yu}},
  \bibinfo {author} {\bibfnamefont {A.~V.}\ \bibnamefont {Kretinin}}, \bibinfo
  {author} {\bibfnamefont {F.}~\bibnamefont {Withers}}, \bibinfo {author}
  {\bibfnamefont {Y.}~\bibnamefont {Cao}}, \bibinfo {author} {\bibfnamefont
  {A.}~\bibnamefont {Mishchenko}}, \bibinfo {author} {\bibfnamefont {I.~V.}\
  \bibnamefont {Grigorieva}}, \bibinfo {author} {\bibfnamefont {K.~S.}\
  \bibnamefont {Novoselov}}, \bibinfo {author} {\bibfnamefont {L.~S.}\
  \bibnamefont {Levitov}}, \ and\ \bibinfo {author} {\bibfnamefont {A.~K.}\
  \bibnamefont {Geim}},\ }\href {\doibase 10.1126/science.1254966} {\bibfield
  {journal} {\bibinfo  {journal} {Science}\ }\textbf {\bibinfo {volume}
  {346}},\ \bibinfo {pages} {448} (\bibinfo {year} {2014})}\BibitemShut
  {NoStop}%
\bibitem [{\citenamefont {Sui}\ \emph {et~al.}(2015)\citenamefont {Sui},
  \citenamefont {Chen}, \citenamefont {Ma}, \citenamefont {Shan}, \citenamefont
  {Tian}, \citenamefont {Watanabe}, \citenamefont {Taniguchi}, \citenamefont
  {Jin}, \citenamefont {Yao}, \citenamefont {Xiao},\ and\ \citenamefont
  {Zhang}}]{sui2015}%
  \BibitemOpen
  \bibfield  {author} {\bibinfo {author} {\bibfnamefont {M.}~\bibnamefont
  {Sui}}, \bibinfo {author} {\bibfnamefont {G.}~\bibnamefont {Chen}}, \bibinfo
  {author} {\bibfnamefont {L.}~\bibnamefont {Ma}}, \bibinfo {author}
  {\bibfnamefont {W.-Y.}\ \bibnamefont {Shan}}, \bibinfo {author}
  {\bibfnamefont {D.}~\bibnamefont {Tian}}, \bibinfo {author} {\bibfnamefont
  {K.}~\bibnamefont {Watanabe}}, \bibinfo {author} {\bibfnamefont
  {T.}~\bibnamefont {Taniguchi}}, \bibinfo {author} {\bibfnamefont
  {X.}~\bibnamefont {Jin}}, \bibinfo {author} {\bibfnamefont {W.}~\bibnamefont
  {Yao}}, \bibinfo {author} {\bibfnamefont {D.}~\bibnamefont {Xiao}}, \ and\
  \bibinfo {author} {\bibfnamefont {Y.}~\bibnamefont {Zhang}},\ }\href
  {\doibase 10.1038/NPHYS3485} {\bibfield  {journal} {\bibinfo  {journal}
  {Nature Phys.}\ }\textbf {\bibinfo {volume} {11}},\ \bibinfo {pages} {1027}
  (\bibinfo {year} {2015})}\BibitemShut {NoStop}%
\bibitem [{\citenamefont {Shimazaki}\ \emph {et~al.}(2015)\citenamefont
  {Shimazaki}, \citenamefont {Yamamoto}, \citenamefont {Borzenets},
  \citenamefont {Watanabe}, \citenamefont {Taniguchi},\ and\ \citenamefont
  {Tarucha}}]{shimazaki2015}%
  \BibitemOpen
  \bibfield  {author} {\bibinfo {author} {\bibfnamefont {Y.}~\bibnamefont
  {Shimazaki}}, \bibinfo {author} {\bibfnamefont {M.}~\bibnamefont {Yamamoto}},
  \bibinfo {author} {\bibfnamefont {I.~V.}\ \bibnamefont {Borzenets}}, \bibinfo
  {author} {\bibfnamefont {K.}~\bibnamefont {Watanabe}}, \bibinfo {author}
  {\bibfnamefont {T.}~\bibnamefont {Taniguchi}}, \ and\ \bibinfo {author}
  {\bibfnamefont {S.}~\bibnamefont {Tarucha}},\ }\href {\doibase
  10.1038/NPHYS3551} {\bibfield  {journal} {\bibinfo  {journal} {Nature Phys.}\
  }\textbf {\bibinfo {volume} {11}},\ \bibinfo {pages} {1032} (\bibinfo {year}
  {2015})}\BibitemShut {NoStop}%
\bibitem [{\citenamefont {{Wu}}\ \emph {et~al.}(2019)\citenamefont {{Wu}},
  \citenamefont {{Zhou}}, \citenamefont {{Liu}}, \citenamefont {{Lin}},
  \citenamefont {{Han}}, \citenamefont {{An}}, \citenamefont {{Wang}},
  \citenamefont {{Xu}}, \citenamefont {{Long}}, \citenamefont {{Cheng}},
  \citenamefont {{Tuen Law}}, \citenamefont {{Zhang}},\ and\ \citenamefont
  {{Wang}}}]{wu2019}%
  \BibitemOpen
  \bibfield  {author} {\bibinfo {author} {\bibfnamefont {Z.}~\bibnamefont
  {{Wu}}}, \bibinfo {author} {\bibfnamefont {B.~T.}\ \bibnamefont {{Zhou}}},
  \bibinfo {author} {\bibfnamefont {G.-B.}\ \bibnamefont {{Liu}}}, \bibinfo
  {author} {\bibfnamefont {J.}~\bibnamefont {{Lin}}}, \bibinfo {author}
  {\bibfnamefont {T.}~\bibnamefont {{Han}}}, \bibinfo {author} {\bibfnamefont
  {L.}~\bibnamefont {{An}}}, \bibinfo {author} {\bibfnamefont {Y.}~\bibnamefont
  {{Wang}}}, \bibinfo {author} {\bibfnamefont {S.}~\bibnamefont {{Xu}}},
  \bibinfo {author} {\bibfnamefont {G.}~\bibnamefont {{Long}}}, \bibinfo
  {author} {\bibfnamefont {C.}~\bibnamefont {{Cheng}}}, \bibinfo {author}
  {\bibfnamefont {K.}~\bibnamefont {{Tuen Law}}}, \bibinfo {author}
  {\bibfnamefont {F.}~\bibnamefont {{Zhang}}}, \ and\ \bibinfo {author}
  {\bibfnamefont {N.}~\bibnamefont {{Wang}}},\ }\href {\doibase
  10.1038/s41467-019-08629-9} {\bibfield  {journal} {\bibinfo  {journal} {Nat.
  Commun.}\ }\textbf {\bibinfo {volume} {10}},\ \bibinfo {pages} {611}
  (\bibinfo {year} {2019})}\BibitemShut {NoStop}%
\bibitem [{\citenamefont {{Hung}}\ \emph {et~al.}(2019)\citenamefont {{Hung}},
  \citenamefont {{Camsari}}, \citenamefont {{Zhang}}, \citenamefont
  {{Upadhyaya}},\ and\ \citenamefont {{Chen}}}]{hung2019}%
  \BibitemOpen
  \bibfield  {author} {\bibinfo {author} {\bibfnamefont {T.~Y.~T.}\
  \bibnamefont {{Hung}}}, \bibinfo {author} {\bibfnamefont {K.~Y.}\
  \bibnamefont {{Camsari}}}, \bibinfo {author} {\bibfnamefont {S.}~\bibnamefont
  {{Zhang}}}, \bibinfo {author} {\bibfnamefont {P.}~\bibnamefont
  {{Upadhyaya}}}, \ and\ \bibinfo {author} {\bibfnamefont {Z.}~\bibnamefont
  {{Chen}}},\ }\href {https://advances.sciencemag.org/content/5/4/eaau6478}
  {\bibfield  {journal} {\bibinfo  {journal} {Science Advances}\ }\textbf
  {\bibinfo {volume} {5}},\ \bibinfo {pages} {eaau6478} (\bibinfo {year}
  {2019})}\BibitemShut {NoStop}%
\bibitem [{\citenamefont {Shan}\ and\ \citenamefont {Xiao}(2019)}]{shan2019}%
  \BibitemOpen
  \bibfield  {author} {\bibinfo {author} {\bibfnamefont {W.-Y.}\ \bibnamefont
  {Shan}}\ and\ \bibinfo {author} {\bibfnamefont {D.}~\bibnamefont {Xiao}},\
  }\href {\doibase 10.1103/PhysRevB.99.205416} {\bibfield  {journal} {\bibinfo
  {journal} {Phys. Rev. B}\ }\textbf {\bibinfo {volume} {99}},\ \bibinfo
  {pages} {205416} (\bibinfo {year} {2019})}\BibitemShut {NoStop}%
\bibitem [{\citenamefont {Kalameitsev}\ \emph {et~al.}(2019)\citenamefont
  {Kalameitsev}, \citenamefont {Kovalev},\ and\ \citenamefont
  {Savenko}}]{kalameitsev2019}%
  \BibitemOpen
  \bibfield  {author} {\bibinfo {author} {\bibfnamefont {A.~V.}\ \bibnamefont
  {Kalameitsev}}, \bibinfo {author} {\bibfnamefont {V.~M.}\ \bibnamefont
  {Kovalev}}, \ and\ \bibinfo {author} {\bibfnamefont {I.~G.}\ \bibnamefont
  {Savenko}},\ }\href {\doibase 10.1103/PhysRevLett.122.256801} {\bibfield
  {journal} {\bibinfo  {journal} {Phys. Rev. Lett.}\ }\textbf {\bibinfo
  {volume} {122}},\ \bibinfo {pages} {256801} (\bibinfo {year}
  {2019})}\BibitemShut {NoStop}%
\bibitem [{\citenamefont {Vozmediano}\ \emph {et~al.}(2010)\citenamefont
  {Vozmediano}, \citenamefont {Katsnelson},\ and\ \citenamefont
  {Guinea}}]{vozmediano2010}%
  \BibitemOpen
  \bibfield  {author} {\bibinfo {author} {\bibfnamefont {M.~A.~H.}\
  \bibnamefont {Vozmediano}}, \bibinfo {author} {\bibfnamefont {M.~I.}\
  \bibnamefont {Katsnelson}}, \ and\ \bibinfo {author} {\bibfnamefont
  {F.}~\bibnamefont {Guinea}},\ }\href {\doibase 10.1016/j.physrep.2010.07.003}
  {\bibfield  {journal} {\bibinfo  {journal} {Phys. Rep.}\ }\textbf {\bibinfo
  {volume} {496}},\ \bibinfo {pages} {109} (\bibinfo {year}
  {2010})}\BibitemShut {NoStop}%
\bibitem [{\citenamefont {Guinea}\ \emph {et~al.}(2010)\citenamefont {Guinea},
  \citenamefont {Katsnelson},\ and\ \citenamefont {Geim}}]{guinea2010}%
  \BibitemOpen
  \bibfield  {author} {\bibinfo {author} {\bibfnamefont {F.}~\bibnamefont
  {Guinea}}, \bibinfo {author} {\bibfnamefont {M.~I.}\ \bibnamefont
  {Katsnelson}}, \ and\ \bibinfo {author} {\bibfnamefont {A.~K.}\ \bibnamefont
  {Geim}},\ }\href {\doibase 10.1038/nphys1420} {\bibfield  {journal} {\bibinfo
   {journal} {Nature Physics}\ }\textbf {\bibinfo {volume} {6}},\ \bibinfo
  {pages} {30} (\bibinfo {year} {2010})}\BibitemShut {NoStop}%
\bibitem [{\citenamefont {Levy}\ \emph {et~al.}(2010)\citenamefont {Levy},
  \citenamefont {Burke}, \citenamefont {Meaker}, \citenamefont {Panlasigui},
  \citenamefont {Zettl}, \citenamefont {Guinea}, \citenamefont {Castro~Neto},\
  and\ \citenamefont {Crommie}}]{levy2010}%
  \BibitemOpen
  \bibfield  {author} {\bibinfo {author} {\bibfnamefont {N.}~\bibnamefont
  {Levy}}, \bibinfo {author} {\bibfnamefont {S.~A.}\ \bibnamefont {Burke}},
  \bibinfo {author} {\bibfnamefont {K.~L.}\ \bibnamefont {Meaker}}, \bibinfo
  {author} {\bibfnamefont {M.}~\bibnamefont {Panlasigui}}, \bibinfo {author}
  {\bibfnamefont {A.}~\bibnamefont {Zettl}}, \bibinfo {author} {\bibfnamefont
  {F.}~\bibnamefont {Guinea}}, \bibinfo {author} {\bibfnamefont {A.~H.}\
  \bibnamefont {Castro~Neto}}, \ and\ \bibinfo {author} {\bibfnamefont {M.~F.}\
  \bibnamefont {Crommie}},\ }\href {\doibase 10.1126/science.1191700}
  {\bibfield  {journal} {\bibinfo  {journal} {Science}\ }\textbf {\bibinfo
  {volume} {329}},\ \bibinfo {pages} {544} (\bibinfo {year}
  {2010})}\BibitemShut {NoStop}%
\bibitem [{\citenamefont {Cazalilla}\ \emph {et~al.}(2014)\citenamefont
  {Cazalilla}, \citenamefont {Ochoa},\ and\ \citenamefont
  {Guinea}}]{cazalilla2014}%
  \BibitemOpen
  \bibfield  {author} {\bibinfo {author} {\bibfnamefont {M.~A.}\ \bibnamefont
  {Cazalilla}}, \bibinfo {author} {\bibfnamefont {H.}~\bibnamefont {Ochoa}}, \
  and\ \bibinfo {author} {\bibfnamefont {F.}~\bibnamefont {Guinea}},\ }\href
  {\doibase 10.1103/PhysRevLett.113.077201} {\bibfield  {journal} {\bibinfo
  {journal} {Phys. Rev. Lett.}\ }\textbf {\bibinfo {volume} {113}},\ \bibinfo
  {pages} {077201} (\bibinfo {year} {2014})}\BibitemShut {NoStop}%
\bibitem [{\citenamefont {Yang}\ \emph {et~al.}(2011)\citenamefont {Yang},
  \citenamefont {Pan}, \citenamefont {Yao},\ and\ \citenamefont
  {Niu}}]{yang2011}%
  \BibitemOpen
  \bibfield  {author} {\bibinfo {author} {\bibfnamefont {S.~A.}\ \bibnamefont
  {Yang}}, \bibinfo {author} {\bibfnamefont {H.}~\bibnamefont {Pan}}, \bibinfo
  {author} {\bibfnamefont {Y.}~\bibnamefont {Yao}}, \ and\ \bibinfo {author}
  {\bibfnamefont {Q.}~\bibnamefont {Niu}},\ }\href {\doibase
  10.1103/PhysRevB.83.125122} {\bibfield  {journal} {\bibinfo  {journal} {Phys.
  Rev. B}\ }\textbf {\bibinfo {volume} {83}},\ \bibinfo {pages} {125122}
  (\bibinfo {year} {2011})}\BibitemShut {NoStop}%
\bibitem [{\citenamefont {Goerbig}\ \emph {et~al.}(2008)\citenamefont
  {Goerbig}, \citenamefont {Fuchs}, \citenamefont {Montambaux},\ and\
  \citenamefont {Pi\'echon}}]{goerbig2008}%
  \BibitemOpen
  \bibfield  {author} {\bibinfo {author} {\bibfnamefont {M.~O.}\ \bibnamefont
  {Goerbig}}, \bibinfo {author} {\bibfnamefont {J.-N.}\ \bibnamefont {Fuchs}},
  \bibinfo {author} {\bibfnamefont {G.}~\bibnamefont {Montambaux}}, \ and\
  \bibinfo {author} {\bibfnamefont {F.}~\bibnamefont {Pi\'echon}},\ }\href
  {\doibase 10.1103/PhysRevB.78.045415} {\bibfield  {journal} {\bibinfo
  {journal} {Phys. Rev. B}\ }\textbf {\bibinfo {volume} {78}},\ \bibinfo
  {pages} {045415} (\bibinfo {year} {2008})}\BibitemShut {NoStop}%
\bibitem [{\citenamefont {Choi}\ \emph {et~al.}(2010)\citenamefont {Choi},
  \citenamefont {Jhi},\ and\ \citenamefont {Son}}]{choi2010}%
  \BibitemOpen
  \bibfield  {author} {\bibinfo {author} {\bibfnamefont {S.-M.}\ \bibnamefont
  {Choi}}, \bibinfo {author} {\bibfnamefont {S.-H.}\ \bibnamefont {Jhi}}, \
  and\ \bibinfo {author} {\bibfnamefont {Y.-W.}\ \bibnamefont {Son}},\ }\href
  {\doibase 10.1103/PhysRevB.81.081407} {\bibfield  {journal} {\bibinfo
  {journal} {Phys. Rev. B}\ }\textbf {\bibinfo {volume} {81}},\ \bibinfo
  {pages} {081407} (\bibinfo {year} {2010})}\BibitemShut {NoStop}%
\bibitem [{\citenamefont {Kobayashi}\ \emph {et~al.}(2007)\citenamefont
  {Kobayashi}, \citenamefont {Katayama}, \citenamefont {Suzumura},\ and\
  \citenamefont {Fukuyama}}]{kobayashi2007}%
  \BibitemOpen
  \bibfield  {author} {\bibinfo {author} {\bibfnamefont {A.}~\bibnamefont
  {Kobayashi}}, \bibinfo {author} {\bibfnamefont {S.}~\bibnamefont {Katayama}},
  \bibinfo {author} {\bibfnamefont {Y.}~\bibnamefont {Suzumura}}, \ and\
  \bibinfo {author} {\bibfnamefont {H.}~\bibnamefont {Fukuyama}},\ }\href
  {\doibase 10.1143/JPSJ.76.034711} {\bibfield  {journal} {\bibinfo  {journal}
  {J. Phys. Soc. Jpn.}\ }\textbf {\bibinfo {volume} {76}},\ \bibinfo {pages}
  {034711} (\bibinfo {year} {2007})}\BibitemShut {NoStop}%
\bibitem [{\citenamefont {Winkler}\ and\ \citenamefont
  {Z\"ulicke}(2010)}]{winkler2010}%
  \BibitemOpen
  \bibfield  {author} {\bibinfo {author} {\bibfnamefont {R.}~\bibnamefont
  {Winkler}}\ and\ \bibinfo {author} {\bibfnamefont {U.}~\bibnamefont
  {Z\"ulicke}},\ }\href {\doibase 10.1103/PhysRevB.82.245313} {\bibfield
  {journal} {\bibinfo  {journal} {Phys. Rev. B}\ }\textbf {\bibinfo {volume}
  {82}},\ \bibinfo {pages} {245313} (\bibinfo {year} {2010})}\BibitemShut
  {NoStop}%
\bibitem [{\citenamefont {de~Juan}\ \emph {et~al.}(2012)\citenamefont
  {de~Juan}, \citenamefont {Sturla},\ and\ \citenamefont
  {Vozmediano}}]{dejuan2012}%
  \BibitemOpen
  \bibfield  {author} {\bibinfo {author} {\bibfnamefont {F.}~\bibnamefont
  {de~Juan}}, \bibinfo {author} {\bibfnamefont {M.}~\bibnamefont {Sturla}}, \
  and\ \bibinfo {author} {\bibfnamefont {M.~A.~H.}\ \bibnamefont
  {Vozmediano}},\ }\href {\doibase 10.1103/PhysRevLett.108.227205} {\bibfield
  {journal} {\bibinfo  {journal} {Phys. Rev. Lett.}\ }\textbf {\bibinfo
  {volume} {108}},\ \bibinfo {pages} {227205} (\bibinfo {year}
  {2012})}\BibitemShut {NoStop}%
\bibitem [{\citenamefont {de~Juan}\ \emph {et~al.}(2013)\citenamefont
  {de~Juan}, \citenamefont {Ma\~nes},\ and\ \citenamefont
  {Vozmediano}}]{dejuan2013}%
  \BibitemOpen
  \bibfield  {author} {\bibinfo {author} {\bibfnamefont {F.}~\bibnamefont
  {de~Juan}}, \bibinfo {author} {\bibfnamefont {J.~L.}\ \bibnamefont
  {Ma\~nes}}, \ and\ \bibinfo {author} {\bibfnamefont {M.~A.~H.}\ \bibnamefont
  {Vozmediano}},\ }\href {\doibase 10.1103/PhysRevB.87.165131} {\bibfield
  {journal} {\bibinfo  {journal} {Phys. Rev. B}\ }\textbf {\bibinfo {volume}
  {87}},\ \bibinfo {pages} {165131} (\bibinfo {year} {2013})}\BibitemShut
  {NoStop}%
\bibitem [{\citenamefont {Ma\~nes}\ \emph {et~al.}(2013)\citenamefont
  {Ma\~nes}, \citenamefont {de~Juan}, \citenamefont {Sturla},\ and\
  \citenamefont {Vozmediano}}]{manes2013}%
  \BibitemOpen
  \bibfield  {author} {\bibinfo {author} {\bibfnamefont {J.~L.}\ \bibnamefont
  {Ma\~nes}}, \bibinfo {author} {\bibfnamefont {F.}~\bibnamefont {de~Juan}},
  \bibinfo {author} {\bibfnamefont {M.}~\bibnamefont {Sturla}}, \ and\ \bibinfo
  {author} {\bibfnamefont {M.~A.~H.}\ \bibnamefont {Vozmediano}},\ }\href
  {\doibase 10.1103/PhysRevB.88.155405} {\bibfield  {journal} {\bibinfo
  {journal} {Phys. Rev. B}\ }\textbf {\bibinfo {volume} {88}},\ \bibinfo
  {pages} {155405} (\bibinfo {year} {2013})}\BibitemShut {NoStop}%
\bibitem [{lan()}]{landau1959}%
  \BibitemOpen
  \href@noop {} {}\bibinfo {note} {L. D. Landau and E. M. Lifschitz,
  \emph{Theory of Elasticity}, Pergamon Press, Oxford, 1959.}\BibitemShut
  {Stop}%
\bibitem [{\citenamefont {Dyson}(1962)}]{dyson1962}%
  \BibitemOpen
  \bibfield  {author} {\bibinfo {author} {\bibfnamefont {F.~J.}\ \bibnamefont
  {Dyson}},\ }\href {\doibase 10.1063/1.1703773} {\bibfield  {journal}
  {\bibinfo  {journal} {J. Math. Phys. (NY)}\ }\textbf {\bibinfo {volume}
  {3}},\ \bibinfo {pages} {140} (\bibinfo {year} {1962})}\BibitemShut {NoStop}%
\bibitem [{\citenamefont {Shan}\ \emph {et~al.}(2012)\citenamefont {Shan},
  \citenamefont {Lu},\ and\ \citenamefont {Shen}}]{shan2012}%
  \BibitemOpen
  \bibfield  {author} {\bibinfo {author} {\bibfnamefont {W.-Y.}\ \bibnamefont
  {Shan}}, \bibinfo {author} {\bibfnamefont {H.-Z.}\ \bibnamefont {Lu}}, \ and\
  \bibinfo {author} {\bibfnamefont {S.-Q.}\ \bibnamefont {Shen}},\ }\href
  {\doibase 10.1103/PhysRevB.86.125303} {\bibfield  {journal} {\bibinfo
  {journal} {Phys. Rev. B}\ }\textbf {\bibinfo {volume} {86}},\ \bibinfo
  {pages} {125303} (\bibinfo {year} {2012})}\BibitemShut {NoStop}%
\bibitem [{\citenamefont {Luican}\ \emph {et~al.}(2011)\citenamefont {Luican},
  \citenamefont {Li},\ and\ \citenamefont {Andrei}}]{luican2011}%
  \BibitemOpen
  \bibfield  {author} {\bibinfo {author} {\bibfnamefont {A.}~\bibnamefont
  {Luican}}, \bibinfo {author} {\bibfnamefont {G.}~\bibnamefont {Li}}, \ and\
  \bibinfo {author} {\bibfnamefont {E.~Y.}\ \bibnamefont {Andrei}},\ }\href
  {\doibase 10.1103/PhysRevB.83.041405} {\bibfield  {journal} {\bibinfo
  {journal} {Phys. Rev. B}\ }\textbf {\bibinfo {volume} {83}},\ \bibinfo
  {pages} {041405} (\bibinfo {year} {2011})}\BibitemShut {NoStop}%
\bibitem [{\citenamefont {Yan}\ \emph {et~al.}(2013)\citenamefont {Yan},
  \citenamefont {Chu}, \citenamefont {Yan}, \citenamefont {Liu}, \citenamefont
  {Meng}, \citenamefont {Yang}, \citenamefont {Fan}, \citenamefont {Wang},
  \citenamefont {Dou}, \citenamefont {Zhang}, \citenamefont {Liu},
  \citenamefont {Nie},\ and\ \citenamefont {He}}]{yan2013}%
  \BibitemOpen
  \bibfield  {author} {\bibinfo {author} {\bibfnamefont {H.}~\bibnamefont
  {Yan}}, \bibinfo {author} {\bibfnamefont {Z.-D.}\ \bibnamefont {Chu}},
  \bibinfo {author} {\bibfnamefont {W.}~\bibnamefont {Yan}}, \bibinfo {author}
  {\bibfnamefont {M.}~\bibnamefont {Liu}}, \bibinfo {author} {\bibfnamefont
  {L.}~\bibnamefont {Meng}}, \bibinfo {author} {\bibfnamefont {M.}~\bibnamefont
  {Yang}}, \bibinfo {author} {\bibfnamefont {Y.}~\bibnamefont {Fan}}, \bibinfo
  {author} {\bibfnamefont {J.}~\bibnamefont {Wang}}, \bibinfo {author}
  {\bibfnamefont {R.-F.}\ \bibnamefont {Dou}}, \bibinfo {author} {\bibfnamefont
  {Y.}~\bibnamefont {Zhang}}, \bibinfo {author} {\bibfnamefont
  {Z.}~\bibnamefont {Liu}}, \bibinfo {author} {\bibfnamefont {J.-C.}\
  \bibnamefont {Nie}}, \ and\ \bibinfo {author} {\bibfnamefont
  {L.}~\bibnamefont {He}},\ }\href {\doibase 10.1103/PhysRevB.87.075405}
  {\bibfield  {journal} {\bibinfo  {journal} {Phys. Rev. B}\ }\textbf {\bibinfo
  {volume} {87}},\ \bibinfo {pages} {075405} (\bibinfo {year}
  {2013})}\BibitemShut {NoStop}%
\bibitem [{\citenamefont {Jang}\ \emph {et~al.}(2014)\citenamefont {Jang},
  \citenamefont {Kim}, \citenamefont {Shin}, \citenamefont {Wang},
  \citenamefont {Jang}, \citenamefont {Kim}, \citenamefont {Lee}, \citenamefont
  {Kim}, \citenamefont {Song},\ and\ \citenamefont {Kahng}}]{jang2014}%
  \BibitemOpen
  \bibfield  {author} {\bibinfo {author} {\bibfnamefont {W.-J.}\ \bibnamefont
  {Jang}}, \bibinfo {author} {\bibfnamefont {H.}~\bibnamefont {Kim}}, \bibinfo
  {author} {\bibfnamefont {Y.-R.}\ \bibnamefont {Shin}}, \bibinfo {author}
  {\bibfnamefont {M.}~\bibnamefont {Wang}}, \bibinfo {author} {\bibfnamefont
  {S.~K.}\ \bibnamefont {Jang}}, \bibinfo {author} {\bibfnamefont
  {M.}~\bibnamefont {Kim}}, \bibinfo {author} {\bibfnamefont {S.}~\bibnamefont
  {Lee}}, \bibinfo {author} {\bibfnamefont {S.-W.}\ \bibnamefont {Kim}},
  \bibinfo {author} {\bibfnamefont {Y.~J.}\ \bibnamefont {Song}}, \ and\
  \bibinfo {author} {\bibfnamefont {S.-J.}\ \bibnamefont {Kahng}},\ }\href
  {\doibase doi.org/10.1016/j.carbon.2014.03.015} {\bibfield  {journal}
  {\bibinfo  {journal} {Carbon}\ }\textbf {\bibinfo {volume} {74}},\ \bibinfo
  {pages} {139} (\bibinfo {year} {2014})}\BibitemShut {NoStop}%
\bibitem [{\citenamefont {Sinitsyn}\ \emph {et~al.}(2006)\citenamefont
  {Sinitsyn}, \citenamefont {Niu},\ and\ \citenamefont
  {MacDonald}}]{sinitsyn2006}%
  \BibitemOpen
  \bibfield  {author} {\bibinfo {author} {\bibfnamefont {N.~A.}\ \bibnamefont
  {Sinitsyn}}, \bibinfo {author} {\bibfnamefont {Q.}~\bibnamefont {Niu}}, \
  and\ \bibinfo {author} {\bibfnamefont {A.~H.}\ \bibnamefont {MacDonald}},\
  }\href {\doibase 10.1103/PhysRevB.73.075318} {\bibfield  {journal} {\bibinfo
  {journal} {Phys. Rev. B}\ }\textbf {\bibinfo {volume} {73}},\ \bibinfo
  {pages} {075318} (\bibinfo {year} {2006})}\BibitemShut {NoStop}%
\bibitem [{sup()}]{supple}%
  \BibitemOpen
  \href@noop {} {}\bibinfo {note} {See Supplementary Material for calculation
  details.}\BibitemShut {Stop}%
\bibitem [{\citenamefont {Chakraborty}\ \emph {et~al.}(2012)\citenamefont
  {Chakraborty}, \citenamefont {Bera}, \citenamefont {Muthu}, \citenamefont
  {Bhowmick}, \citenamefont {Waghmare},\ and\ \citenamefont
  {Sood}}]{chakraborty2012}%
  \BibitemOpen
  \bibfield  {author} {\bibinfo {author} {\bibfnamefont {B.}~\bibnamefont
  {Chakraborty}}, \bibinfo {author} {\bibfnamefont {A.}~\bibnamefont {Bera}},
  \bibinfo {author} {\bibfnamefont {D.~V.~S.}\ \bibnamefont {Muthu}}, \bibinfo
  {author} {\bibfnamefont {S.}~\bibnamefont {Bhowmick}}, \bibinfo {author}
  {\bibfnamefont {U.~V.}\ \bibnamefont {Waghmare}}, \ and\ \bibinfo {author}
  {\bibfnamefont {A.~K.}\ \bibnamefont {Sood}},\ }\href {\doibase
  10.1103/PhysRevB.85.161403} {\bibfield  {journal} {\bibinfo  {journal} {Phys.
  Rev. B}\ }\textbf {\bibinfo {volume} {85}},\ \bibinfo {pages} {161403}
  (\bibinfo {year} {2012})}\BibitemShut {NoStop}%
\bibitem [{\citenamefont {Kumaravadivel}\ \emph {et~al.}(2019)\citenamefont
  {Kumaravadivel}, \citenamefont {Greenaway}, \citenamefont {Perello},
  \citenamefont {Berdyugin}, \citenamefont {Birkbeck}, \citenamefont {Wengraf},
  \citenamefont {Liu}, \citenamefont {Edgar}, \citenamefont {Geim},
  \citenamefont {Eaves},\ and\ \citenamefont
  {Krishna~Kumar}}]{kumaravadivel2019}%
  \BibitemOpen
  \bibfield  {author} {\bibinfo {author} {\bibfnamefont {P.}~\bibnamefont
  {Kumaravadivel}}, \bibinfo {author} {\bibfnamefont {M.~T.}\ \bibnamefont
  {Greenaway}}, \bibinfo {author} {\bibfnamefont {D.}~\bibnamefont {Perello}},
  \bibinfo {author} {\bibfnamefont {A.}~\bibnamefont {Berdyugin}}, \bibinfo
  {author} {\bibfnamefont {J.}~\bibnamefont {Birkbeck}}, \bibinfo {author}
  {\bibfnamefont {J.}~\bibnamefont {Wengraf}}, \bibinfo {author} {\bibfnamefont
  {S.}~\bibnamefont {Liu}}, \bibinfo {author} {\bibfnamefont {J.~H.}\
  \bibnamefont {Edgar}}, \bibinfo {author} {\bibfnamefont {A.~K.}\ \bibnamefont
  {Geim}}, \bibinfo {author} {\bibfnamefont {L.}~\bibnamefont {Eaves}}, \ and\
  \bibinfo {author} {\bibfnamefont {R.}~\bibnamefont {Krishna~Kumar}},\ }\href
  {\doibase 10.1038/s41467-019-11379-3} {\bibfield  {journal} {\bibinfo
  {journal} {Nature Communications}\ }\textbf {\bibinfo {volume} {10}},\
  \bibinfo {pages} {3334} (\bibinfo {year} {2019})}\BibitemShut {NoStop}%
\bibitem [{\citenamefont {Zhang}\ and\ \citenamefont {Niu}(2015)}]{zhang2015}%
  \BibitemOpen
  \bibfield  {author} {\bibinfo {author} {\bibfnamefont {L.}~\bibnamefont
  {Zhang}}\ and\ \bibinfo {author} {\bibfnamefont {Q.}~\bibnamefont {Niu}},\
  }\href {\doibase 10.1103/PhysRevLett.115.115502} {\bibfield  {journal}
  {\bibinfo  {journal} {Phys. Rev. Lett.}\ }\textbf {\bibinfo {volume} {115}},\
  \bibinfo {pages} {115502} (\bibinfo {year} {2015})}\BibitemShut {NoStop}%
\bibitem [{\citenamefont {Ishizuka}\ and\ \citenamefont
  {Nagaosa}(2018)}]{ishizuka2018}%
  \BibitemOpen
  \bibfield  {author} {\bibinfo {author} {\bibfnamefont {H.}~\bibnamefont
  {Ishizuka}}\ and\ \bibinfo {author} {\bibfnamefont {N.}~\bibnamefont
  {Nagaosa}},\ }\href
  {https://advances.sciencemag.org/content/4/2/eaap9962/tab-pdf} {\bibfield
  {journal} {\bibinfo  {journal} {Science Advances}\ }\textbf {\bibinfo
  {volume} {4}},\ \bibinfo {pages} {eaap9962} (\bibinfo {year}
  {2018})}\BibitemShut {NoStop}%
\bibitem [{\citenamefont {Kato}\ and\ \citenamefont
  {Ishizuka}(2019)}]{kato2019}%
  \BibitemOpen
  \bibfield  {author} {\bibinfo {author} {\bibfnamefont {Y.}~\bibnamefont
  {Kato}}\ and\ \bibinfo {author} {\bibfnamefont {H.}~\bibnamefont
  {Ishizuka}},\ }\href {\doibase 10.1103/PhysRevApplied.12.021001} {\bibfield
  {journal} {\bibinfo  {journal} {Phys. Rev. Applied}\ }\textbf {\bibinfo
  {volume} {12}},\ \bibinfo {pages} {021001} (\bibinfo {year}
  {2019})}\BibitemShut {NoStop}%
\bibitem [{\citenamefont {Okamoto}\ \emph {et~al.}(2019)\citenamefont
  {Okamoto}, \citenamefont {Egami},\ and\ \citenamefont
  {Nagaosa}}]{okamoto2019}%
  \BibitemOpen
  \bibfield  {author} {\bibinfo {author} {\bibfnamefont {S.}~\bibnamefont
  {Okamoto}}, \bibinfo {author} {\bibfnamefont {T.}~\bibnamefont {Egami}}, \
  and\ \bibinfo {author} {\bibfnamefont {N.}~\bibnamefont {Nagaosa}},\ }\href
  {\doibase 10.1103/PhysRevLett.123.196603} {\bibfield  {journal} {\bibinfo
  {journal} {Phys. Rev. Lett.}\ }\textbf {\bibinfo {volume} {123}},\ \bibinfo
  {pages} {196603} (\bibinfo {year} {2019})}\BibitemShut {NoStop}%
\bibitem [{\citenamefont {Rostami}\ \emph {et~al.}(2015)\citenamefont
  {Rostami}, \citenamefont {Rold\'an}, \citenamefont {Cappelluti},
  \citenamefont {Asgari},\ and\ \citenamefont {Guinea}}]{rostami2015}%
  \BibitemOpen
  \bibfield  {author} {\bibinfo {author} {\bibfnamefont {H.}~\bibnamefont
  {Rostami}}, \bibinfo {author} {\bibfnamefont {R.}~\bibnamefont {Rold\'an}},
  \bibinfo {author} {\bibfnamefont {E.}~\bibnamefont {Cappelluti}}, \bibinfo
  {author} {\bibfnamefont {R.}~\bibnamefont {Asgari}}, \ and\ \bibinfo {author}
  {\bibfnamefont {F.}~\bibnamefont {Guinea}},\ }\href {\doibase
  10.1103/PhysRevB.92.195402} {\bibfield  {journal} {\bibinfo  {journal} {Phys.
  Rev. B}\ }\textbf {\bibinfo {volume} {92}},\ \bibinfo {pages} {195402}
  (\bibinfo {year} {2015})}\BibitemShut {NoStop}%
\bibitem [{\citenamefont {Ochoa}(2019)}]{ochoa2019}%
  \BibitemOpen
  \bibfield  {author} {\bibinfo {author} {\bibfnamefont {H.}~\bibnamefont
  {Ochoa}},\ }\href {\doibase 10.1103/PhysRevB.100.155426} {\bibfield
  {journal} {\bibinfo  {journal} {Phys. Rev. B}\ }\textbf {\bibinfo {volume}
  {100}},\ \bibinfo {pages} {155426} (\bibinfo {year} {2019})}\BibitemShut
  {NoStop}%
\bibitem [{\citenamefont {Lian}\ \emph {et~al.}(2019)\citenamefont {Lian},
  \citenamefont {Wang},\ and\ \citenamefont {Bernevig}}]{lian2019}%
  \BibitemOpen
  \bibfield  {author} {\bibinfo {author} {\bibfnamefont {B.}~\bibnamefont
  {Lian}}, \bibinfo {author} {\bibfnamefont {Z.}~\bibnamefont {Wang}}, \ and\
  \bibinfo {author} {\bibfnamefont {B.~A.}\ \bibnamefont {Bernevig}},\ }\href
  {\doibase 10.1103/PhysRevLett.122.257002} {\bibfield  {journal} {\bibinfo
  {journal} {Phys. Rev. Lett.}\ }\textbf {\bibinfo {volume} {122}},\ \bibinfo
  {pages} {257002} (\bibinfo {year} {2019})}\BibitemShut {NoStop}%
\end{thebibliography}
\end{document}